\documentclass[english,letterpaper,aps,prl,twocolumn,superscriptaddress]{revtex4-1}
\usepackage{amssymb}
\usepackage{amsbsy}
\usepackage{amsmath}
\usepackage{graphicx}
\usepackage{graphics}
\usepackage{setspace}
\usepackage{array}
\usepackage{color}
\usepackage{fontenc}
\usepackage{textcomp}
\usepackage{rotating}
\usepackage{bm}
\usepackage{bbm}
\usepackage{lmodern}
\usepackage{dcolumn}
\usepackage{wasysym}
\usepackage{marvosym}
\usepackage{subfloat}
\usepackage{color}
\usepackage{subfigure}
\usepackage{tikz}
\usepackage{pdfpages}

\makeatletter
\AtBeginDocument{\let\LS@rot\@undefined}
\makeatother

\makeatletter
\@ifundefined{textcolor}{}
{%
 \definecolor{BLACK}{gray}{0}
 \definecolor{WHITE}{gray}{1}
 \definecolor{RED}{rgb}{1,0,0}
 \definecolor{GREEN}{rgb}{0,1,0}
 \definecolor{BLUE}{rgb}{0,0,1}
 \definecolor{CYAN}{cmyk}{1,0,0,0}
 \definecolor{MAGENTA}{cmyk}{0,1,0,0}
 \definecolor{YELLOW}{cmyk}{0,0,1,0}
}

\newlength{\textwidthm}

   \let\x=\xi

\def\bpm{\begin{pmatrix}}
\def\epm{\end{pmatrix}}
\def\be{\begin{equation}}
\def\ee{\end{equation}}
\def\bea{\begin{eqnarray}}
\def\eea{\end{eqnarray}}
\def\ba{\begin{array}}
\def\ea{\end{array}}

\newcommand{\mathsym}[1]{{}}
\newcommand{\unicode}[1]{{}}

\begin{document}

\title{Quantum Geometry and Topology of Bulk Plasmons in Weyl Metals}

\author{Hong-Yi Xie}
\email{monkxhy@gmail.com}
\affiliation{Department of Physics and Astronomy, Center for Quantum Research
and Technology, University of Oklahoma, Norman, OK 73069, USA}

\author{Peter Abbamonte}
\affiliation{Department of Physics, University of Illinois, Urbana, IL, 61801}

\author{Bruno Uchoa}
\email{uchoa@ou.edu}
\affiliation{Department of Physics and Astronomy, Center for Quantum Research
and Technology, University of Oklahoma, Norman, OK 73069, USA} 

\date{\today}

\begin{abstract}
We address the quantum geometric structure of plasmons in Fermi surfaces enclosing a topological charge. 
We demonstrate that Weyl fermion plasmons  have monopole structure, are topological and have a finite vorticity $\zeta=2\mathsf{C}_{\text{w}}$, where $\mathsf{C}_{\text{w}}$
is the Chern number of the Fermi surface enclosing the Weyl point. 
We show that these plasmons selectively couple to light linearly polarized along the plasmon effective dipole moment $\mathbf{d}$, which has quantum geometric origin and points along the direction of the plasmon center of mass momentum $\hat{\mathbf{Q}}$. 
We suggest that Weyl metal topological plasmons have distinctive optical properties compared to conventional plasmons. 
\end{abstract}

\maketitle

\emph{Introduction.}--- Plasmons are collective particle-hole excitations of the Fermi surface in metals. 
Although they are commonly described semiclassically, they have an internal quantum structure that results from the quantum geometry of the Hilbert space. 
In 2D topological insulators, the quantum geometry has been shown to produce a plasmon Berry dipole, which was predicted to cause electronic skew scattering~\cite{Cao}. 
Bosonic collective modes in interacting electron systems may inherit the quantum geometry and the topology of the electron bands. 
Quantum geometric effects can stabilize flat-band superconductivity~\cite{Peotta,Julku,Torma,Jiang}, can lead to the emergence of topological excitons in flat Chern bands~\cite{Xie, Wu3, Xie2, Kwan} and to the development of topological Cooper pairing in Weyl metals~\cite{Li}. 
Magnons of flat-band ferromagnets were recently proposed to have a quantum geometric dipole \cite{Chen}. 

Despite having unique properties when coupled to surface plasmons and Fermi arcs~\cite{Song,Andolina,Adinehvand,Chen-1,Gordin,Ghosh,Lu}, and being proposed as indicators~\cite{Zhou} of the chiral anomaly~\cite{Son,Parameswaran,Andreev,Xiong,Huang,Moll}, plasmons in Weyl metals have not been recognized as being intrinsically topological objects themselves. 
Also, in all known Weyl metals~\cite{Armitage,Liu,Morali,Belopolski,Okamura}, non-topological Fermi surfaces dominate nearly all the spectral weight, making the experimental observation of Weyl fermion plasmons challenging. In this Letter, we address the nature of plasmons in topologically nontrivial Fermi surfaces with Chern number $\mathsf{C}_{\text{W}}$. We demonstrate that plasmons in three dimensional (3D) Fermi surfaces enclosing a monopole charge encode both the quantum geometry and the topology of the fermions. 

\begin{figure}[b]
\includegraphics[scale=0.25]{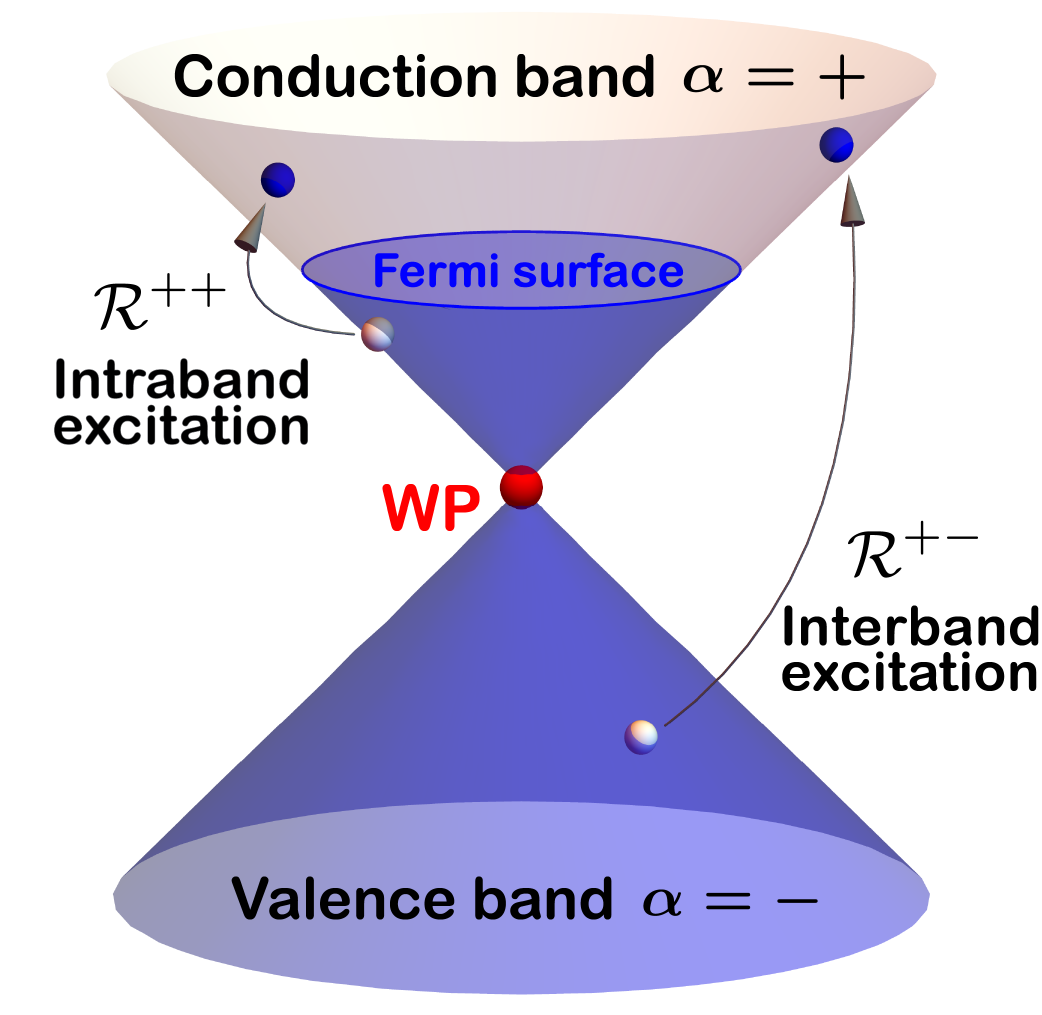}
\caption{Energy bands near a Weyl point (WP) enclosed by a Fermi surface. 
Plasmon processes involve both intraband and interband particle-hole
transitions, which contribute to the plasmon envelope function $\mathcal{R}_{\mathbf{q}}^{\alpha\beta}(\mathbf{Q})$.
The interband component $\mathcal{R}^{+-}$ ($\mathcal{R}^{-+}$) has monopole structure and has a finite vorticity in the relative
momentum coordinates $\mathbf{q}$. } \label{fig1}

\end{figure}

To be concrete, we consider a Weyl metal node with two energy branches, $\alpha=\pm$, accounting for the conduction and valence bands. Plasmons can be generically expressed as superpositions of particle-hole excitations with center of mass (COM) momentum $\mathbf{Q}$ around a Fermi surface~\cite{Cao,Sawada}. In the electron basis, the ket state of a plasmon
has the form 
\begin{equation}
|\mathbf{Q}\rangle=\sum_{\mathbf{k},\alpha,\beta}\mathcal{R}_{\mathbf{k}}^{\alpha\beta}(\mathbf{Q})|\alpha,\mathbf{k}+\mathbf{Q}/2\rangle|\beta,\mathbf{k}-\mathbf{Q}/2\rangle^{\ast},\label{eq:Q}
\end{equation}
where $|\alpha,\mathbf{p}\rangle$ describes an electron with momentum $\mathbf{p}$ in band $\alpha$, $\mathbf{k}$ is the relative momentum
coordinates of the electron-pair state, and $\mathcal{R}_{\mathbf{k}}^{\alpha\beta}(\mathbf{Q})$
are the components of the envelope function of the plasmon. The envelope function is a four-component spinor $\hat{\mathcal{R}}=(\mathcal{R}^{++},\mathcal{R}^{+-},\mathcal{R}^{-+},\mathcal{R}^{--})^{T}$ that incorporates both intraband $(\alpha=\beta)$ and interband ($\alpha=-\beta$)
particle-hole excitations, as illustrated in Fig.~\ref{fig1}. 

Plasmons require intraband particle-hole transitions near the Fermi
surface to exist \cite{Pines,Kotov}. We show that the interband components
of the envelope function, which have no analog in Fermi liquids, have
monopole structure, are topological and have a finite vorticity in
the relative momentum coordinates equal to twice the helicity of the
Weyl fermions, $\zeta=2\mathsf{C}_{\text{W}}\in2\mathbb{Z}$. We predict
that those plasmons couple to linearly polarized light $\mathbf{\boldsymbol{\mathcal{E}}}(t)$
when 
\begin{equation}
\boldsymbol{\mathcal{E}}(t)\cdot\mathbf{d}\neq0,\qquad\mathbf{d}\propto\hat{\mathbf{Q}}.\label{eq:e}
\end{equation}
Here $\mathbf{d}$ is the plasmon effective dipole moment, which has quantum
geometric origin and points in the same direction of the plasmon COM
momentum $\hat{\mathbf{Q}}= \mathbf{Q}/|\mathbf{Q}|$. We suggest that Weyl metal topological plasmons
have unique properties that may permit their observation for the first time. 

\emph{Model.}--- We study the plasmon excitations in a single-valley
of a Weyl metal described by the Hamiltonian 
\begin{equation}
\mathcal{H}=\sum_{\mathbf{k}}\Psi_{\mathbf{k}}^{\dagger}(v\hat{\boldsymbol{\sigma}}\cdot\mathbf{k}-\varepsilon_{F})\Psi_{\mathbf{k}}+\frac{1}{2}\sum_{\mathbf{Q}}v(\mathbf{Q}):\hat{n}_{\mathbf{Q}}\hat{n}_{\mathbf{-Q}}:,
\end{equation}
where $\Psi_{\mathbf{k}}=(\psi_{1,\mathbf{k}},\psi_{2,\mathbf{k}})$
is a two-component fermionic spinor operator with momentum $\mathbf{k}$,
written either in the spin basis or else in the orbital basis of a
spin polarized system, with  $\sum_{\mathbf{k}} \equiv \int\! d^3\mathbf{k}/(2\pi)^3$.   $\hat{\boldsymbol{\sigma}}=(\sigma_{x},\sigma_{y},\sigma_{z})$
is a vector of Pauli matrices, $v$ is the Fermi velocity (we set $\hbar \to 1$)  and $\varepsilon_{F}$
is the Fermi energy away from the Weyl point.
The second term is the normal ordered Coulomb interaction, $v(\mathbf{Q})=4\pi e^{2}/(\epsilon |\mathbf{Q}|^{2})$,
expressed in terms of density fluctuation operators in the same basis, $\hat{n}_{\mathbf{Q}}=\sum_{\mathbf{k}}\Psi_{\mathbf{k}+\mathbf{Q}/2}^{\dagger}\Psi_{\mathbf{k}-\mathbf{Q}/2}$. 

We apply an unitary transformation $\hat{U}(\mathbf{k})$ that diagonalizes the
kinetic part of the Hamiltonian. Rewriting the fermion operators in the energy basis
$c_{\alpha,\mathbf{k}}=\sum_{a=1,2}U_{\alpha a}^{\ast}(\mathbf{k})\psi_{a,\mathbf{k}}$,
the Hamiltonian has the form
\begin{align}
\mathcal{H}_c= & \sum_{\mathbf{k},\alpha=\pm}\varepsilon_{\alpha,\mathbf{k}}c_{\alpha,\mathbf{k}}^{\dagger}c_{\alpha,\mathbf{k}}
+ \frac{1}{2} \!\!\sum_{\mathbf{q},\mathbf{q}^{\prime},\alpha_{1}\ldots\alpha_{4}} \!\!  v(\mathbf{Q}) \,S_{\mathbf{q}+\frac{\mathbf{Q}}{2},\mathbf{q}-\frac{\mathbf{Q}}{2}}^{\alpha_{1}\alpha_{2}}      
  \nonumber \\
   &\qquad \qquad \qquad \times   S_{\mathbf{q}^{\prime}+\frac{\mathbf{Q}}{2},\mathbf{q}^{\prime}-\frac{\mathbf{Q}}{2}}^{\ast,\alpha_{3}\alpha_{4}}   :P_{\mathbf{q},\mathbf{Q}}^{\alpha_{1}\alpha_{2}}P_{\mathbf{q}^{\prime},-\mathbf{Q}}^{\alpha_{4}\alpha_{3}}:,
\end{align}
where $\varepsilon_{\alpha,\mathbf{k}}=\alpha|\mathbf{k}|-\varepsilon_{F}$ is the Weyl dispersion and $P_{\mathbf{q},\mathbf{Q}}^{\alpha\beta} \equiv c_{\alpha,\mathbf{q}+\mathbf{\frac{Q}{2}}}^{\dagger}c_{\beta,\mathbf{q}-\frac{\mathbf{Q}}{2}}$ is the electron-hole pair operator. The form factor matrix
\begin{equation}
S_{\mathbf{k}_{1},\mathbf{k}_{2}}^{\alpha\beta}=\sum_{a=1,2}U_{a \alpha}^{*}(\mathbf{k}_{1})U_{a \beta}(\mathbf{k}_{2}), \label{eq:S}
\end{equation}
defined in band space, encodes the quantum geometry of the Weyl fermions. 
 Expansion of the form factor matrix up to second order around zero COM momentum $\mathbf{Q}$   gives ~\cite{SM} 
\begin{align} 
\hat{S}_{\mathbf{q}+\frac{\mathbf{Q}}{2},\mathbf{q}-\frac{\mathbf{Q}}{2}} \approx \hat{\mathbb{I}} - i Q_i \hat{A}_{i,\mathbf{q}} - \frac{1}{4} Q_i Q_j \mathrm{Tr}(\hat{A}_{i,\mathbf{q}} \hat{A}_{j,\mathbf{q}}) \hat{\mathbb{I}}, \label{S-exp}
\end{align}
where $\hat{\mathbb{I}}$ is identity matrix, $i,j=x,y,z$ and $\hat{\mathbf{A}}_{\mathbf{q}}= -i\hat{U}^\dagger(\mathbf{q}) \nabla_{\mathbf{q}} \hat{U}(\mathbf{q})$ is the Berry connection tensor of the Weyl fermions, which is traceless and Hermitian. In the quadratic term, $\mathrm{Tr}(\hat{A}_{i,\mathbf{q}} \hat{A}_{j,\mathbf{q}}) = g_{ij,\mathbf{q}} + A_{i,\mathbf{q}}^{++}A_{j,\mathbf{q}}^{++}$, where $g_{ij,\mathbf{q}} = \mathrm{Re}(A_{i,\mathbf{q}}^{+-}A_{j,\mathbf{q}}^{-+})$ is the quantum metric tensor of either the conduction or valence bands. 

\emph{Plasmon topology}.--- Plasmons are generically described as
poles of the retarded correlation function of pair operators, 
\begin{equation}
\Theta_{R,\mathbf{q},\mathbf{q}^{\prime}}^{\alpha_{1}\alpha_{2};\alpha_{3}\alpha_{4}}(\mathbf{Q},t)\equiv-i\theta(t)\langle[P_{\mathbf{q},-\mathbf{Q}}^{\alpha_{1}\alpha_{2}}(t), \, P_{\mathbf{q}^{\prime},\mathbf{Q}}^{\alpha_{4}\alpha_{3}}(0)]\rangle,\label{eq:Pi1}
\end{equation}
where $\langle\ldots\rangle$ implies  ensemble average weighted by the thermal equilibrium density matrix of the system. Unlike the conventional
charge susceptibility, which is defined as a correlation function
of density fluctuation operators, this quantity is not averaged over momentum
and has information about electron-hole envelope function. The plasmon dispersion $\omega_{\mathbf{Q}}$
and the envelope function $\mathcal{R}_{\mathbf{q}}^{\alpha\beta}(\mathbf{Q})$
are determined by the zeroes of the inverse correlation function $\hat{\Theta}_{R}^{-1}(\mathbf{Q},\omega)=\int_{-\infty}^{\infty}\text{d}t\,\text{e}^{i\omega t}\hat{\Theta}_{R}^{-1}(\mathbf{Q},t)$,
which forms the Kernel of the Bethe-Salpeter equation at the RPA level,
\begin{equation}
\sum_{\alpha_{3}\alpha_{4},\mathbf{q}^{\prime}}[\Theta_{R}^{-1}(\mathbf{Q},\omega_{\mathbf{Q}})]_{\mathbf{q},\mathbf{q}^{\prime}}^{\alpha_{1}\alpha_{2};\alpha_{3}\alpha_{4}}\,\mathcal{R}_{\mathbf{q}^{\prime}}^{\alpha_{3}\alpha_{4}}(\mathbf{Q})=0,\label{eq:BS}
\end{equation}
where \begin{widetext}
\begin{equation}
[\Theta_{R}^{-1}(\mathbf{Q},\omega)]_{\mathbf{q},\mathbf{q}^{\prime}}^{\alpha_{1}\alpha_{2};\alpha_{3}\alpha_{4}}=\frac{\omega_{+}+\varepsilon_{\alpha_{2},\mathbf{q}-\frac{\mathbf{Q}}{2}}-\varepsilon_{\alpha_{1},\mathbf{q}+\frac{\mathbf{Q}}{2}}}{f_{\alpha_{1},\mathbf{q}+\frac{\mathbf{Q}}{2}}-f_{\alpha_{2},\mathbf{q}-\frac{\mathbf{Q}}{2}}}\delta_{\mathbf{q},\mathbf{q}^{\prime}}\delta_{\alpha_{1}\alpha_{3}}\delta_{\alpha_{2},\alpha_{4}}+v(\mathbf{Q})
S_{\mathbf{q}+\frac{\mathbf{Q}}{2},\mathbf{q}-\frac{\mathbf{Q}}{2}}^{\alpha_{1}\alpha_{2}}
S_{\mathbf{q}^{\prime}+\frac{\mathbf{Q}}{2},\mathbf{q}^{\prime}-\frac{\mathbf{Q}}{2}}^{*,\alpha_{3}\alpha_{4}}.\label{eq:Chi2}
\end{equation}
\end{widetext} 
Here $f_{\alpha,\mathbf{k}}\equiv f_{F}(\varepsilon_{\alpha,\mathbf{k}})$ is the Fermi distribution, $\omega_{+}=\omega+i0^{+}$, and 
$\delta_{\mathbf{q},\mathbf{q}^{\prime}} \equiv (2\pi)^3\delta(\mathbf{q}-\mathbf{q}^{\prime})$. 

\begin{figure*}[t]
\begin{centering}
\includegraphics[scale=0.26]{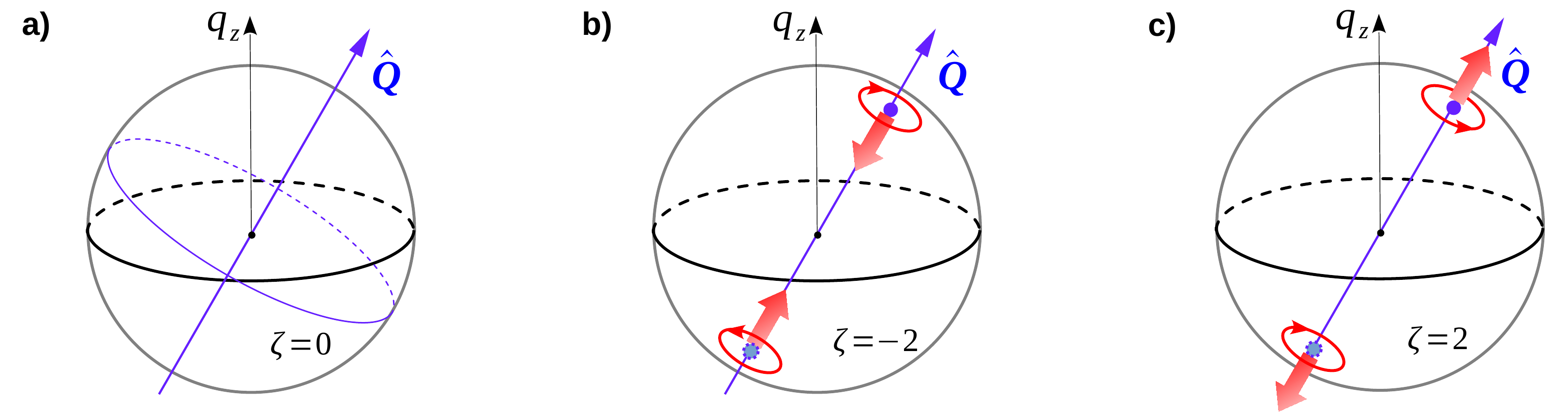}
\par\end{centering}
\caption{{\small Illustration of the topological structure of the envelope functions
$\mathcal{R}_{\mathbf{q}}^{\alpha\beta}(\mathbf{Q}\to0)$. The sphere
(gray line) is defined by the polar and azimuthal angles of the relative
momentum $q$. The equator is the branch cut of the $q_{c} \neq 0$ MSH.
The COM momentum axis $\hat{\mathbf{Q}}$ is depicted by the blue
arrow. a) Intraband components $\mathcal{R}_{\mathbf{q}}^{\alpha\alpha}$
($\alpha=\pm$) exhibit a nodal plane (blue circle) orthogonal to
the $\hat{\mathbf{Q}}$ axis, with no vortex present ($\zeta=0$).
b) Interband component $\mathcal{R}_{\mathbf{q}}^{+-}$ exhibits a
nodal line along the $\hat{\mathbf{Q}}$ axis with total vorticity
$\zeta=-2$. Two vortices with vorticity $-1$ (red arrows) form around
the $\hat{\mathbf{Q}}$ axis. c) The interband component $\mathcal{R}_{\mathbf{q}}^{-+}$
exhibits a nodal line along the $\hat{\mathbf{Q}}$ axis, with two
vortices with vorticity $+1$ (red arrows) forming around the nodal
line ($\zeta=2$). }} \label{fig2}
\end{figure*}

The plasmon dispersion is determined via the Sherman-Morrison reduction of the Bethe Salpeter equation~\cite{SM2}. This procedure is equivalent to finding the zeroes of the real part of the dynamical dielectric function $\epsilon(\mathbf{Q},\omega)=1-v(\mathbf{Q})\Pi_{R}(\mathbf{Q},\omega)$,
where $\Pi_{R}$ is the standard retarded RPA polarization bubble
\cite{Zhou,Kotov,Hosur}
\begin{align}
\Pi^{R}(\mathbf{Q},\omega) & =\frac{1}{2}\sum_{\alpha\beta,\mathbf{k}}\left(1+\alpha\beta\hat{\mathbf{k}}\cdot\widehat{\mathbf{k}+\mathbf{Q}}\right)\label{eq:Pi}\nonumber \\
 & \qquad\qquad\times\frac{f_{\alpha,\mathbf{k}}-f_{\beta,\mathbf{k}+\mathbf{Q}}}{\omega_{+}+\varepsilon_{\beta,\mathbf{k}+\mathbf{Q}}-\varepsilon_{\alpha,\mathbf{k}}},
\end{align}
with $\hat{\mathbf{k}}=\mathbf{k}/k$ a unit vector. The solution
of the plasmon frequency in the $\mathbf{Q}\to 0$ limit including interband processes, obtained
through the analytical evaluation of Eq.~(\ref{eq:Pi}), is
\begin{equation} 
\omega_{\mathbf{Q}\to 0}\approx \frac{\alpha_0  |\varepsilon_F|}{\sqrt{1+\frac{\alpha_0^2}{2} \ln \frac{\Lambda}{k_F} }},\label{eq:omega0}
\end{equation}
where $\Lambda$ is the ultraviolet cut-off, $\alpha_{0}=\sqrt{e^{2}/(6\pi^{2}\epsilon v)}\lesssim1$
is the effective fine structure constant and $k_{F}=|\varepsilon_{F}|/v$ is the Fermi momentum. The logarithmic correction due to interband processes red shifts the plasmon frequency \cite{SM3} and is parametrically small in large Fermi surfaces. In
small Fermi surfaces, it introduces a cut-off dependent typical value
for the Fermi momentum $k_F^\ast$, estimated by $\ln(\Lambda/k_{F}^{*})\sim 2/\alpha_{0}^{2}$. 
When $k_F < k_F^\ast$, the Bethe-Salpeter equation  breaks down and  plasmon formation is strongly suppressed. 
As shown below, this constraint sets an upper bound for the magnitude of the quantum geometry induced plasmon effective dipole moment.

The exact solution of the Bethe-Salpeter Eq. (\ref{eq:BS}) is
\begin{equation}
\mathcal{R}_{\eta,\mathbf{q}}^{\alpha\beta}(\mathbf{Q})=\mathcal{N}(Q)\frac{f_{\alpha,\mathbf{q}+\frac{\mathbf{Q}}{2}}-f_{\beta,\mathbf{q}-\frac{\mathbf{Q}}{2}}}{\eta\omega_{\mathbf{Q}}+\varepsilon_{\beta,\mathbf{q}-\frac{\mathbf{Q}}{2}}-\varepsilon_{\alpha,\mathbf{q}+\frac{\mathbf{Q}}{2}}}S_{\mathbf{q}+\frac{\mathbf{Q}}{2},\mathbf{q}-\frac{\mathbf{Q}}{2}}^{\alpha\beta},\label{eq:R}
\end{equation}
where $\mathcal{N}(Q)$ is an SO$(3)$ invariant normalization factor,
$\mathcal{N}(Q\to0)\to\mathcal{N}_{0}/Q$, with $\mathcal{N}_{0}$
a constant and $\eta=\pm1$ labels the two plasmon branches \cite{SM4}.   

The topology of the envelope function components can be classified
according to the theory of spin-weighted spherical harmonics \cite{Gelfand,Newman}.
Under a $\varphi$ rotation about the $z$-axis in the pseudospin
space, the single-particle basis transforms as $\hat{U}_{\mathbf{k}}\to\hat{U}_{\mathbf{k}}\,\text{e}^{i\sigma_{3}\varphi/2}$.
The envelope function transforms as
\begin{equation} 
\mathcal{R}_{\eta,\mathbf{q}}^{\alpha\beta}(\mathbf{Q})\to\mathcal{R}_{\eta,\mathbf{q}}^{\alpha\beta}(\mathbf{Q})\text{e}^{i(\beta-\alpha)\varphi/2},\label{eq:varphi}
\end{equation}
with a pseudospin weight $s_{\alpha\beta}=(\beta-\alpha)/2$. In the representation of monopole spherical harmonics (MSH) \cite{Li,Wu,Wu2,Haldane}, this corresponds to the `vorticity'' $\zeta_{\alpha\beta}=2s_{\alpha\beta}=\beta-\alpha$~\cite{Dray}. Clearly the interband components have vorticity $\zeta_{\alpha\bar{\alpha}}=-2 \alpha$ (we define $\bar{\alpha} \equiv -\alpha$), whereas the intraband ones have vorticity $\zeta_{\alpha\alpha}=0$. 

In zero COM momentum limit $\mathbf{Q} \to 0$, the envelope function (\ref{eq:R}) can be written in a more explicit form due to the SO(3) symmetry of the normalization factor. The intraband components read
\begin{align}
& \mathcal{R}_{\eta,\mathbf{q}}^{\alpha\alpha}(\mathbf{Q}\to0)= R_{\eta}^{\alpha\alpha}(q)\,\hat{\mathbf{q}}\cdot\hat{\mathbf{Q}}\nonumber \\
& = \frac{4\pi}{3}R_{\eta}^{\alpha\alpha}(q) \sum_{|m|\leq1}(-1)^{m}Y_{1,-m}(\Theta,\Phi) \, Y_{1,m}(\theta,\phi),\label{eq:Rintra}
\end{align}
where $(\theta,\phi)$ and $(\Theta,\Phi)$ are the spherical-coordinate
angles of the relative and COM momentum coordinates, $\hat{\mathbf{q}}$
and $\hat{\mathbf{Q}}$, respectively, $Y_{l,m}$ is a spherical harmonic,
and $R_{\eta}^{\alpha\alpha}(q)=\mathcal{N}_{0}v/(\eta\omega_{0})\partial_{\varepsilon}f_{F}(\varepsilon_{\alpha,q})$.
These components exhibit a nodal plane orthogonal to the $\hat{\mathbf{Q}}$
axis ($\hat{\mathbf{q}}\cdot\hat{\mathbf{Q}}=0$) and describe a $p$-wave
state, with angular momentum $l=1$, as illustrated in Fig.~\ref{fig2}(a). 

The interband components of the envelope function (\ref{eq:R}), on the other hand, depend explicitly on the interband components of
the Berry connection tensor defined below Eq.~\eqref{S-exp}, which has a $U(1)$ gauge symmetry. In the Wu-Yang gauge \cite{Wu,Wu2},
in which the single-particle wavefunctions are analytical at northern
and southern poles of the sphere but have a branch cut in the form
of a phase discontinuity at the equator,
\begin{align}
& \mathcal{R}_{\eta,\mathbf{q}}^{\alpha\bar{\alpha}}(\mathbf{Q}\to0)=  i R_{\eta}^{\alpha\bar{\alpha}}(q) \,\hat{\mathbf{A}}^{\alpha\bar{\alpha}}\cdot\hat{\mathbf{Q}} \nonumber \\
& = \frac{2\sqrt{2}\pi}{3q}R_{\eta}^{\alpha\bar{\alpha}}(q) \sum_{|m|\leq 1}(-1)^{m}Y_{1,-m}(\Theta,\Phi)\mathcal{Y}_{\bar{\alpha};1,m}(\theta,\phi),\label{eq:Rinter}
\end{align}
where $\mathcal{Y}_{q_{c};l,m}$ is a MSH with monopole charge $q_{c}$ and
\begin{equation}
R_{\eta}^{\alpha\bar{\alpha}}(q)= \mathcal{N}_{0}\frac{f_{\bar{\alpha},q}-f_{\alpha,q}}{\eta\omega_{0}-2 \alpha v q}. \label{eq:R2}
\end{equation}
The component $\mathcal{R}_{\eta,\mathbf{q}}^{\alpha\bar{\alpha}}$ has a nodal line along the $\hat{\mathbf{Q}}$ axis ($\hat{\mathbf{q}}\cdot\hat{\mathbf{Q}}=\pm1)$, with two vortices on the sphere having the same vorticity $\bar{\alpha}$. Therefore, the total vorticity $\zeta= 2 \bar{\alpha} \in \pm 2$ corresponds to a
pair of vortices ($\alpha = -$) or anti-vortices ($\alpha=+$) around an axis oriented along the
COM momentum direction $\hat{\mathbf{Q}}$, as shown respectively in Figs.~\ref{fig2}(b) and (c).

Plasmons with finite vorticity are expected to have distinctive optical properties and selection rules. This feature may permit distinguishing them from conventional plasmons, which form in non-topological Fermi surfaces in Weyl metals. In two dimensions, charge-neutral bosonic collective excitations analogous to chiral
graviton modes with vorticity $\zeta = \pm 2$ have been recently observed in fractional quantum Hall liquids ~\cite{Liang}. 

In spite of their non-trivial topology in the relative momentum coordinates,
Weyl fermion plasmons are gapped bosonic modes in 3D, and hence,
are  topologically trivial in the COM coordinates, where
they have zero Berry connection in the pseudospin basis, $-i\langle\mathbf{Q}|\nabla_{\mathbf{Q}}|\mathbf{Q}\rangle=0$
\cite{SM5}. A non-trivial topological structure in COM momentum could nevertheless still
emerge from the crossing of two or more plasmon branches.

\emph{Optical selection rule}.--- 
An external light field $\boldsymbol{\mathcal{E}}(t)$ couples to the Weyl fermions as 
\begin{equation}
\mathcal{H}_{\mathcal{E}}(t)=- e \boldsymbol{\mathcal{E}}(t) \cdot \sum_{\alpha\beta,\mathbf{k}} \mathbf{A}_\mathbf{k}^{\alpha\beta} c_{\alpha \mathbf{k}}^\dagger c_{\beta \mathbf{k}},
\end{equation} 
with $\hat{\mathbf{A}}_{\mathbf{k}}$ the Berry connection tensor. 
The optical response of the system is determined by the time-dependent single-particle density matrix $\rho_{\mathbf{q},\mathbf{Q}}^{\alpha\beta}(t) \equiv \langle c_{\beta,\mathbf{q}-\mathbf{Q}/2}^\dagger (t) c_{\alpha,\mathbf{q}+\mathbf{Q}/2}(t) \rangle$~\cite{Haug,SM}. For weak light field, the density matrix can be expanded perturbatively in the light-matter coupling, $\rho_{\mathbf{k},\mathbf{Q}}^{\alpha\beta}(t) = \sum_{n \ge 0} \rho_{\mathbf{k},\mathbf{Q}}^{(n),\alpha\beta}(t)$, where $\rho_{\mathbf{k},\mathbf{Q}}^{(n),\alpha\beta}(t) \sim \mathcal{E}^n$. Assuming that the system is initially in equilibrium ($\boldsymbol{\mathcal{E}}=0$), we obtain the RPA equation of motion for the density matrix in linear response \cite{SM6},
\begin{equation}
\sum_{\alpha\beta,\mathbf{q}^{\prime}}\left[\hat{\Theta}_{R}^{-1}(\omega)\right]_{\mathbf{q},\mathbf{q}^{\prime}}^{\alpha_{1}\alpha_{2};\alpha_{3}\alpha_{4}}\rho_{\mathbf{q}^{\prime}}^{(1),\alpha_{3}\alpha_{4}}(\omega)=e\boldsymbol{\mathcal{E}}(\omega)\cdot\mathbf{A}_{\mathbf{q}}^{\alpha_{1}\alpha_{2}}.\label{eq:Source}
\end{equation}
The solution of this inhomogenous equation is
\begin{equation}
\rho_{\mathbf{q}}^{(1),\alpha\beta}(\omega)= \frac{\boldsymbol{\mathcal{E}}(\omega)\cdot\mathbf{d}}{\omega_{+}-\omega_{0}}\mathcal{R}_{\mathbf{q}}^{\alpha\beta}(0) \label{eq:X}.
\end{equation}
The dynamical part of the density matrix in linear response is resonant at the plasmon frequency $\omega_{0}$ and implies that the plasmons are longitudinally polarized. $\mathbf{d}$ is the plasmon effective electric dipole moment, 
\begin{equation}
\mathbf{d}=e \sum_{\alpha\beta,\mathbf{q}}\mathcal{R}_{\mathbf{q}}^{*,\alpha\beta}\mathbf{A}_{\mathbf{q}}^{\alpha\beta}=d_{0}\hat{\mathbf{Q}}, \label{eq:d}
\end{equation}
which has quantum geometric origin and points along the plasmon COM momentum. Using Eqs.~\eqref{eq:Rintra} and \eqref{eq:Rinter}, one can explicitly evaluate the amplitude of the plasmon dipole moment, 
\begin{equation}
d_{0}= \frac{e \mathcal{N}_0}{12 \hbar v \pi^2}  \ln \left( \frac{\Lambda}{k_F} \right) \lesssim \frac{e \mathcal{N}_0}{6 \hbar v \pi^2 \alpha_0^2},\label{eq:d0}
\end{equation}
with the upper bound in the logarithmic dependence of the integral set by the condition of validity of the Bethe-Salpeter equation, outlined below Eq. (\ref{eq:omega0}). 

The total polarization due to the plasmon mode in the presence of
the light source is $\boldsymbol{\mathcal{P}}(\omega)=e\sum_{\alpha\beta,\mathbf{k}}\rho_{\mathbf{k}}^{(1),\beta\alpha}(\omega)\mathbf{A}_{\mathbf{k}}^{\alpha\beta}$
\cite{Haug}. Combining this relation with Eqs.~\eqref{eq:X} and \eqref{eq:d}, one can reexpress the polarization in linear response in the external
electric field as $\mathcal{P}_{i}(\omega)=\sum_{j=x,y,z}\chi_{ij}(\omega)\mathcal{E}_{j}(\omega)$,
where 
\begin{equation}
\chi_{ij}(\omega)= \frac{\theta(\omega)d_{i}^{*}d_{j}}{\omega_{+}-\omega_{0}}-\frac{\theta(-\omega)d_{j}^{*}d_{i}}{\omega_{+}+\omega_{0}}, \label{eq:chi}
\end{equation}
is the susceptibility tensor, which determines the plasmon optical
response. 

Diagonalization of the dyadic tensor $\mathsf{D}_{ij}=d_{i}^{*}d_{j}$
through the eigenvalue equation $\boldsymbol{\mathsf{D}}\cdot\boldsymbol{\mathsf{u}}_{s}=a_{s}\boldsymbol{\mathsf{u}}_{s}$,
with $s=0,1,2$, gives only one non-zero eigenvalue $a_{0}=|d_{0}|^{2}$
with eigenmode $\boldsymbol{\mathsf{u}}_{0}=\hat{\mathbf{d}}^{*}\propto\hat{\mathbf{Q}}$
and two eigenmodes with zero eigenvalue. Hence,  we conclude that plasmons of a Fermi surface containing a monopole charge have a selection rule in which they 
couple with light linearly polarized along the COM momentum $\mathbf{Q}$ and are transparent to light polarized in a perpendicular direction, as anticipated in Eq. (\ref{eq:e}). This contrasts with  the behavior of conventional plasmons, whose dyadic tensor $\mathsf{D}_{ij} \propto \delta_{ij}$, and thus can equally couple to light polarized in any direction.  We note that the optical selection
rules for plasmons in topological Fermi surfaces in 3D can be fundamentally
distinct of the case of charge neutral bosonic modes with finite vorticity
in two dimensions, which tend to couple to circularly polarized light \cite{Lozano}. 


\emph{Experimental observation.}--- Due to their finite electric dipole moment $\mathbf{d}$, plasmons with finite vorticity break inversion symmetry (per Weyl node) and can exhibit second harmonic generation with frequency $2\omega_{\mathbf{Q}\to0}$ in the presence of an external light source resonant with the plasmon frequency $\omega_{\mathbf{Q}\to0}$ \cite{Morimoto}. In a multivalley Weyl metal, Coulomb interaction hybridizes valley-resolved density oscillations, producing collective plasmon eigenmodes whose dipole moments are coherent combinations of those associated with different Weyl nodes. When inversion symmetry relates opposite-chirality valleys, second-order electric-dipole responses are symmetry constrained and may cancel in the bulk. Once inversion symmetry is broken globally \cite{Wu4}, however, this cancellation is no longer protected. The resulting non-degenerate hybridized plasmon modes can contribute to second-harmonic generation as well as DC photocurrent, with nonlinear processes involving both inter-plasmon transitions and transitions between plasmon modes and the electron-hole continuum. The theory of topological plasmon-mediated nonlinear photocurrents in Weyl metals will be addressed in detail elsewhere \cite{Jones}.



\emph{Conclusion.}--- We demonstrated through the specific example of Weyl metals that bulk plasmons of Fermi surfaces enclosing a monopole charge are topological and encode the quantum geometry of the fermion bands. The interband components of the plasmon profile function have monopole structure and have a finite vorticity in the relative momentum coordinates around an axis oriented along the plasmon COM momentum. Those plasmons have  a finite electric dipole moment with quantum geometric origin and selectively couple to light linearly polarized along that axis. Topological plasmons have unique optical properties, which may permit their experimental observation.  We thank M. Mitrano, E. Fradkin and L. England for helpful discussions. BU was supported by NSF grant No. DMR-2529526.



\pagebreak

\makeatletter
\let\cleardoublepage\clearpage
\makeatother

\onecolumngrid

\pagestyle{empty}

\foreach \x in {1,...,8} {%
    \clearpage
    \thispagestyle{empty} 
    
 \hoffset=-1in
    \voffset=-1in
    \oddsidemargin=0pt
    \evensidemargin=0pt
    \topmargin=0pt
    \headheight=0pt
    \headsep=0pt
    \topskip=0pt
    
    \noindent
    \hspace*{-0.71in}%
    \makebox[0pt][l]{%
        \parbox[t][\paperheight]{\paperwidth}{%
            \vspace*{-0.7in}
            \includegraphics[page=\x,width=\paperwidth-0.2in,keepaspectratio=ture]{SM_arX.pdf}%
        }%
    }%
}

\clearpage

\begin{thebibliography}{10}
\bibitem{Cao} J.~Cao, H.~A.~Fertig, and L.~Brey, Quantum internal structure of plasmons, Phys. Rev. Lett. \textbf{127}, 196403 (2021).

\bibitem{Peotta} S.~Peotta and P.~Torma, Superfluidity in topologically nontrivial flat bands, Nat. Commun. \textbf{6}, 8944 (2015).

\bibitem{Julku} A.~Julku, S.~Peotta, T.~I.~Vanhala, D.-H.~Kim, and P.~T\"orm\"a, Geometric origin of superfluidity in the Lieb-lattice flat band, Phys. Rev. Lett. \textbf{117}, 045303 (2016).

\bibitem{Torma} P.~T\"orm\"a, S.~Peotta, and B.~A.~Bernevig, Superfluidity and quantum geometry in twisted multilayer systems, Nat. Rev. Phys. \textbf{4}, 528 (2022).

\bibitem{Jiang} G.~Jiang and Y.~Barlas, Pair density waves from local band geometry, Phys. Rev. Lett. \textbf{131}, 016002 (2023).

\bibitem{Xie} H.-Y.~Xie, P.~Ghaemi, M.~Mitrano, and B.~Uchoa, Theory of topological exciton insulators and condensates in flat Chern bands, Proc. Natl. Acad. Sci. USA \textbf{121}, e2401644121 (2024).

\bibitem{Wu3} F. Wu, T. Lovorn, and A. H. MacDonald, Topological exciton bands in Moire heterojunctions, Phys. Rev. Lett. 118, 147401 (2017).

\bibitem{Xie2} M. Xie, M. Hafezi, and S. Das Sarma, Long-Lived topologi- cal flatband excitons in semiconductor Moire heterostructures: A bosonic Kane-Mele model platform, Phys. Rev. Lett. 133, 136403 (2024).

\bibitem{Kwan} Y. H. Kwan, Z. Wang, G. Wagner, S. H. Simon, S. A. Parameswaran, and Nick Bultinck, Textured exciton insulators, Phys. Rev. B 112, 035129 (2025).

\bibitem{Li} Y.~Li and F.~D.~M.~Haldane, Topological nodal Cooper pairing in doped Weyl metals, Phys. Rev. Lett. \textbf{120}, 067003 (2018).

\bibitem{Chen} L.~Chen, S.~A.~Ghorashi, J.~Cano, and V.~Cr\'epel, Quantum-geometric dipole: a topological boost to flavor ferromagnetism in flat bands, arXiv:2506.22417 [cond-mat.mes-hall]. 

\bibitem{Song} J.~C.~W.~Song and M.~S.~Rudner, Fermi arc plasmons in Weyl semimetals, Phys. Rev. B \textbf{96}, 205443 (2017).

\bibitem{Andolina} G.~M.~Andolina, F.~M.~D.~Pellegrino, F.~H.~L.~Koppens, and M.~Polini, Quantum nonlocal theory of topological Fermi arc plasmons in Weyl semimetals, Phys. Rev. B \textbf{97}, 125431 (2018).

\bibitem{Adinehvand} F.~Adinehvand, Z.~Faraei, T.~Farajollahpour, and S.~A.~Jafari, Sound of Fermi arcs: a linearly dispersing gapless surface plasmon mode in undoped Weyl semimetals, Phys. Rev. B \textbf{100}, 195408 (2019).

\bibitem{Chen-1} Q.~Chen, A.~Ryan~Kutayiah, I.~Oladyshkin, M.~Tokman, and A. Belyanin, Optical properties and electromagnetic modes of Weyl semimetals, Phys. Rev. B \textbf{99} 075137 (2019).

\bibitem{Gordin}E.~V.~Gorbar, V.~A.~Miransky, I.~A.~Shovkovy, and P.~O.~Sukhachov, Hydrodynamics of Fermi arcs: bulk flow and surface collective modes, Phys. Rev. B \textbf{99}, 155120 (2019).

\bibitem{Ghosh}S.~Ghosh and C.~Timm, Dynamical density and spin response of Fermi arcs and their consequences for Weyl semimetals, Phys. Rev. B \textbf{101}, 165402 (2020).

\bibitem{Lu} X.~Lu, D.~K.~Mukherjee, and M.~O.~Goerbig, Surface plasmonics of Weyl semimetals. Phys. Rev. B \textbf{104}, 155103 (2021).

\bibitem{Zhou}J.~Zhou, H.-R.~Chang, and D.~Xiao, Plasmon mode as a detection of the chiral anomaly in Weyl semimetals, Phys. Rev. B \textbf{91}, 035114 (2015).

\bibitem{Son} D.~T.~Son and B.~Z.~Spivak, Chiral anomaly and classical negative magnetoresistance of Weyl metals, Phys. Rev. B \textbf{88}, 104412 (2013).

\bibitem{Parameswaran} S.~A.~Parameswaran, T.~Grover, D.~A.~Abanin, D.~A.~Pesin, and A.~Vishwanath, Probing the chiral anomaly with nonlocal
transport in three-dimensional topological semimetals, Phys. Rev. X \textbf{4}, 031035 (2014).

\bibitem{Andreev} B.~Z.~Spivak and A.~V.~Andreev, Magnetotransport phenomena related to the chiral anomaly in Weyl semimetals, Phys. Rev. B \textbf{93},
085107 (2016).

\bibitem{Xiong}J.~Xiong, S.~K.~Kushwaha, T.~Liang, J.~W.~Krizan, M.~Hirschberger, W.~Wang, R.~J.~Cava, and N.~P.~Ong, Evidence for the chiral anomaly in the Dirac semimetal Na$_{3}$Bi, Science \textbf{350}, 413 (2015).

\bibitem{Huang} X.~Huang, L.~Zhao, Y.~Long, P.~Wang, D.~Chen, Z.~Yang, H.~Liang, M.~Xue, H.~Weng, Z.~Fang, X.~Dai, and G.~Chen, Observation
of the chiral-anomaly-induced negative magnetoresistance in 3D Weyl semimetal TaAs, Phys. Rev. X \textbf{5}, 031023 (2015).

\bibitem{Moll} P.~J.~W.~Moll, N.~L.~Nair, T.~Helm, A.~C.~Potter, I.~Kimchi, A.~Vishwanath, and J.~G.~Analytis, Transport evidence for Fermi-arc-mediated chirality transfer in the Dirac semimetal Cd$_{3}$As$_{2}$, Nature \textbf{535}, 266 (2016).

\bibitem{Armitage}N.~P.~Armitage, E.~J.~Mele, and A.~Vishwanath, Weyl and Dirac semimetals in three-dimensional solids, Rev. Mod. Phys. \textbf{90}, 015001 (2018), and references therein. 

\bibitem{Liu} D.~F.~Liu \emph{et al.}, Magnetic Weyl semimetal phase in a kagome crystal, Science \textbf{365} 1282 (2019). 

\bibitem{Morali} N.~Morali \emph{et al.}, Fermi-arc diversity on surface terminations of the magnetic Weyl semimetal Co$_{3}$Sn$_{2}$S$_{2}$, Science \textbf{365}, 1286 (2019).

\bibitem{Belopolski} I.~Belopolski \emph{et al.}, Discovery of topological Weyl fermion lines and drumhead surface states in a room temperature magnet, Science \textbf{365}, 1278 (2019).

\bibitem{Okamura} Y.~Okamura \emph{et al.}, Giant magneto-optical responses in magnetic Weyl semimetal Co$_{3}$Sn$_{2}$S$_{2}$, Nat. Comm. \textbf{11}, 1 (2020).

\bibitem{Sawada}K.~Sawada, K.~A.~Brueckner, N.~Fukuda, and R.~Brout, Correlation energy of an electron gas at high density: plasma oscillations, Phys. Rev. \textbf{108}, 507 (1957).

\bibitem{Pines}D. Pines, Elementary Excitations in Solids (W. A. Benjamin, 1963).

\bibitem{Kotov}V. N. Kotov, B. Uchoa, V. M. Pereira, F. Guines, and
A. H. Castro Neto, Rev. Mod. Phys. \textbf{84}, 1067 (2012). 

\bibitem{SM} For details on the expansion of the form factor matrix, see supplemental materials.

\bibitem{SM2} For more details on the reduction of the Bethe-Salpeter equation, see supplemental materials. 

\bibitem{Hosur}P. Hosur, S. A. Parameswaran, and A. Vishwanath, Phys.
Rev. Lett. \textbf{108}, 046602 (2012).

\bibitem{SM3} See supplemental materials for the derivation of the plasmon frequency red shift due to interband processes. 

\bibitem{SM4} For details on the normalization constant, see supplemental materials. 

\bibitem{Gelfand}I. M. Gelfand, R. A. Minlos, and G. Cummins, Representations
of the Rotation and Lorentz Groups and Their Applications, (Martino
Fine Books, 2012).

\bibitem{Newman}E. T. Newman and R. Penrose, Note on the Bondi-Metzner-Sachs
Group, J. Math. Phys. 7, 863 (1966).

\bibitem{Wu}T. T. Wu and C. N. Yang, Dirac monopole without strings:
Monopole harmonics, Nucl. Phys. B\textbf{107}, 365 (1976).

\bibitem{Wu2}T. T. Wu and C. N. Yang, Some properties of monopole
harmonics, Phys. Rev. D \textbf{16}, 1018 (1977).

\bibitem{Haldane}F. D. M. Haldane, Fractional Quantization of the
Hall effect: a hierarchy of incompressible quantum fluid states, Phys.
Rev. Lett. \textbf{51}, 605 (1983).

\bibitem{Dray}T. Dray, The relationship between monopole harmonics
and spin- weighted spherical harmonics, J. Math. Phys. 26, 1030 (1984).

\bibitem{Liang}J. Liang, Z. Liu, Z. Yang, Y. Huang, U. Wurstbauer,
C. R. Dean, K. W. West, L. N. Pfeiffer, L. Du, and A. Pinczuk, Nature
\textbf{78}, 628 (2024). 

\bibitem{SM5} See supplemental materials for details on the plasmon Berry connection. 

\bibitem{Haug}H. Haug and S. Koch, Quantum Theory of the Optical
and Electronic Properties of Semiconductors (World Scientific, Singapore,
2004).

\bibitem{SM6} The derivation of the equation of motion for the density matrix in linear response is shown in the supplemental materials. 

\bibitem{Lozano}M. Lozano, H.-Y. Xie , and B. Uchoa, Optical selection
rules of topological excitons in flat bands, Phys. Rev. B \textbf{112},
235417 (2025).

\bibitem{Wu4} L. Wu, S. Patankar, T. Morimoto, N. L. Nair, E. Thewalt, A. Little, J. G. Analytis, J. E. Moore and J. Orenstein, Nat. Phys. {\bf 13}, 350 (2017).

\bibitem{Morimoto}T. Morimoto and N. Nagaosa, Sci. Adv. {\bf 2} e1501524 (2016). 

\bibitem{Jones} L. Jones, H.-Y. Xie, B. Uchoa, unpublished. 

\end{thebibliography}
\end{document}